\documentclass[review]{elsarticle}

\journal{Journal of \LaTeX\ Templates}

\usepackage{multirow}
\usepackage{caption}
\usepackage{subcaption}
\usepackage{xcolor}
\usepackage{diagbox}
\usepackage{longtable}
\usepackage{graphicx}
\usepackage{amsfonts}
\usepackage{hyperref}
\usepackage{amsmath}

\usepackage{array}
\newcolumntype{L}{>{\centering\arraybackslash}m{3cm}}

\begin{document}

\begin{frontmatter}

\title{An application of copulas to OPEC's changing influence on fossil fuel prices}

\author[mymainaddress,mmysecondaryaddress]{Grazian, C.\corref{mycorrespondingauthor}}
\address[mymainaddress]{University of Sydney}
\address[mmysecondaryaddress]{The Data Analytics for Resources \& Environments Centre}
\cortext[mycorrespondingauthor]{Corresponding author}
\ead[url]{https://www.sydney.edu.au/science/about/our-people/academic-staff/clara-grazian.html}
\ead{clara.grazian@sydney.edu.au}

\author[mymainaddress]{McInnes, A.}

\begin{abstract}
This work examines how the dependence structures between energy futures asset prices differ in two periods identified before and after the 2008 global financial crisis. These two periods were characterised by a difference in the number of extraordinary meetings of OPEC countries organised to announce a change of oil production. In the period immediately following the global financial crisis, the decrease in oil prices and oil and gas demand forced OPEC countries to make frequent adjustments to the production of oil, while, since the first quarter of 2010, the recovery led to more regular meetings, with only three organised extraordinary meetings. We propose to use a copula model to study how the dependence structure among energy prices changed among the two periods. The use of copula models allows to introduce flexible and realistic models for the marginal time series; once marginal parameters are estimated, the estimates are used to fit several copula models for all asset combinations. Model selection techniques based on information criteria are implemented to choose the best models both for the univariate asset prices series and for the distribution of co-movements. The changes in the dependence structure of couple of assets are investigated through copula functionals and their uncertainty estimated through a bootstrapping method. We find the strength of dependence between asset combinations considerably differ between the two periods, showing a significant decrease for all the pairs of assets.
\end{abstract}

\end{frontmatter}

\newpage

\section{Introduction}
\label{Introduction}

The 2008 global financial crisis has had strong effects on the global market, both short-term and long-term. In particular, it has affected commodity market prices \citep{chan2011asset}. The crude oil market showed high volatility in the period after the financial crisis: while the crude oil prices constantly increased since the financialization of the commodity market in 2000, they fell from USD 133.88 in June 2008 to USD 39.09 in February 2009 \citep{eseia2021petroleum}, while natural gas fell from USD 12.69 to USD 4.52 in the same period \citep{eseia2021natural}. Figure \ref{fig:figure1b} shows a strong fluctuation during 2008. 

Several studies have tried to investigate the effect of the global financial crisis on the oil market. \cite{martina2011multiscale} used entropy to analyse oil market efficiency and suggested that extreme events should affect the market only short-term; \cite{zhang2017definancialization} studied the correlation of crude oil and natural gas market volatility and identified a breakpoint in 2008 determing the change of volatility; \cite{liu2020commodity} showed that correlation between crude oil futures and sector ETFs increased since the 2008 financial crisis; \cite{joo2020impact} examined the effect of the financial crisis on the oil market in terms of efficiency and long-term equilibrium. Often, studies focus on the immediate effect of the financial crisis, and not on the long-term changes in the oil market; moreover, even when focusing on long-term effects, as \cite{joo2020impact}, the interest is mainly on univariate time series. 

This work aims at modelling the dependence structure among energy indices, including crude oil prices, crude oil derivatives, and natural gas. 
Immediately after the global financial crisis, the Organisation of the Petroleum Exporting Countries (OPEC) organised more frequent meetings, mostly aimed at reducing the oil production and this has been shown to negatively influence the oil market efficiency \cite{jiang2014testing}: the oil market needs time to react to new information. Since 2010, OPEC countries started to meet more regularly. We use the regularity of OPEC announcements as a variable i) associated to the reduction of the volatility experienced after the financial crisis, and ii) representing a factor that \textit{per se} has an effect on the energy market. 

We study the dependence structure of the energy asset prices depending on the regularity of OPEC announcements through copula models \citep{embrechts2001modelling,patton2012review}. There are several advantages in using copulas in this setting; firstly, the dependence structure can be flexibly described and does not have to be necessarily linear or symmetric, and it is possible to model the strenght of the dependnce among extreme values. Secondly, the marginal univariate time series can be modelled in a realistic way, by introducing ARCH effects, asymmetry and heavy tails, in case data show these characteristics, instead of using multivariate models with strong assumptions on both the univariate models and the multivariate dependence, which are often chosen for tractability. For both univariate models and copula models, we consider several choices and apply tools of model selection to identify the model which better represent the data. We show in this work that among the two considered time windows (when OPEC organised many extraordinary meetings and when OPEC organised more regular meetings) the tails of the univariate time series tended to become lighter, with extreme values less likely, and that the dependence among assets, measured as both monotonic dependence and tail dependence, decreased. 

The remainder of the paper is organised as follows. Section \ref{sec:background} introduces the background of energy prices analysis, Section \ref{sec:copulas} described the dataset used in this work, and the univariate and multivariate models that will be used to study the dependence structure among the assets; Section \ref{sec:results} discusses the results, and Section \ref{sec:conclu} concludes the paper. 

\section{Background}
\label{sec:background}

OPEC is the largest and most prominent cartel in the world. It regularly hosts conferences among its thirteen members (Iran, Iraq, Kuwait, Saudi Arabia, Venezuela, Algeria, Angola, Congo, Equatorial Guinea, Gabon, Libya, Nigeria and United Arab Emirates) to agree on oil production policies. Ordinary meetings of OPEC are held twice a year, however extraordinary meetings can also been scheduled. The output of these meetings is usually an announcement, setting country-specific production quotas for its members \citep{secretariat2003opec}. OPEC's ability to influence the energy market has diminished over the years \citep{kaufmann2004does}, however there is still evidence of a significant correlation between the oil prices and variables associated to OPEC (quotas, capacity utilisation, excess of production quotas) \citep{fattouh2005causes, guidi2006effect}. In particular, there is evidence that OPEC announcements can have an impact on market volatility \citep{fattouh2005causes}, for example through the speculation about its decisions; nonetheless, the effect on energy prices time series is not fully understood yet. 

The literature about the effect of OPEC announcements on crude oil markets or energy markets is still limited, and most of it focuses on univariate modelling. \cite{draper1984behavior} studied the impact of scheduled and emergency OPEC announcements on different maturities separately, with some evidence of an effect of scheduled meetings but not of the unscheduled ones. \cite{deaves1992behavior} classified the OPEC announcements into two groups (``good'' and ``bad''), depending on the effect on the returns of oil futures the day after the announcements and use models with ARCH effects to characterize the oil market, showing that ``good'' announcements have a positive effect on the market, while ``bad'' announcements have no significant effect. \cite{wirl2004impact} used a more specific classifications of the announcements with a similar methology. \cite{loutia2016opec} investigated the effect of OPEC production decisions, by classifying them into three categories (increase, cut, maintain), on both WTI and Brent crude oil prices by using separate EGARCH models. \cite{guidi2006effect} analysed the effect of OPEC production changes on stock returns in the United States and in the United Kingdom separately, and showed that reduction of the production seems to have a higher impact. \cite{demirer2010behavior} represents an attempt to analyse together spot and futures oil markets. All these works focus their interest on univariate time series; notably \cite{klein2018trends} studied the linear correlation between crude oil prices at different resolutions, showing that OPEC meetings have little impact on long-term price trends, but they can have an effect on the short-term trends for several days. In particular, negative trends tend to increase the volatility of crude oil prices, with respect to positive trends. 

Our work is different from this literature because i) we consider the regularity of OPEC announcements, ii) we model the returns separately through GARCH models specific to each time series, and iii) we combine the marginal models into a multivariate model through a copula function, in order to analyse the effect of the global finciancial crisis represented by the regularity of OPEC announcements on the dependence structure of energy-related returns. 

Given the well-known limitations of linear correlation to describe dependence among financial returns \citep{embrechts2002correlation}, there is an extensive literature applying copulas to financial markets: see \cite{patton2006modelling} and \cite{ning2010dependence}, among others, for analyses on the relationship between financial and exchange rate markets, and \cite{reboredo2011crude} for an application in the setting of crude oil prices.  
An important task when using copula models to study the dependence among asset prices is the choice of the particular copula function. In the early 2000's the Gaussian copula was a popular tool to model financial data in a flexible way, where individual time series of prices could follow several different marginal distributions. However, it is now recognised that Gaussian copulas, characterised by null tail dependence, can ignore the possibility of occurrence of extreme events \citep{malevergne2003testing, mackenzie2014formula}. Alternative functions to the Gaussian copula are the Student-$t$ copula, which is still an elliptical copula, or any copula function in the class of Archimedean copulas. 

In the setting of energy market, \cite{lu2014portfolio} used copulas with GARCH-type marginals to estimate the Value-at-Risk of crude oil prices and natural gas futures portfolios. \cite{koirala2015energy} used copula models to measure the dependence between energy prices, including crude oil, gasoline and natural gas, and agricultural commodities, using a mixture of Clayton and Gumbel copulas to capture both lower and upper tail depdendence. \cite{ji2018uncertainties} studied the impact of uncertainties on energy prices through copula models, showing that there exists a negative dependence between energy returns and changes in the uncertainty. 
We are contributing to this literature by studying the effect of the regularity of OPEC announcements, as associated to the volatility induced by the global financial crisis, to study the changes in the dependence structure of energy prices, in terms of monotonic dependence, tail dependence, and distance from independence. 


\section{Methods and Models}
\label{sec:copulas}

\subsection{The data}
\label{sec:data}

We consider four assets to represent the energy market: daily prices of West Texas Intermediate (WTI, here indicated as CL) crude oil (originating from the Cushing Oil Field in Oklahoma) and Brent crude oil (originating from the Brent oilfields in the North Sea, here indicated as BZ), Reformulated Gasoline Blendstock for Oxygen Blending (RB), and natural gas (NG). OPEC produced oil is generally priced according to the Brent pricing benchmark, while North American produced oil typically uses the WTI pricing benchmark. RB is considered as a derivative of an OPEC controlled product. Finally, natural gas is considered as an energy asset which should not be influenced by OPEC. 

\begin{figure}[!h]
   \centering
   \begin{subfigure}{0.99\textwidth}
      \centering
\includegraphics[width=6cm,height=5cm]{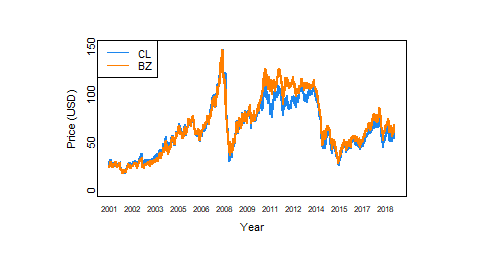}
\hspace{-0.4cm}
\includegraphics[width=6cm,height=5cm]{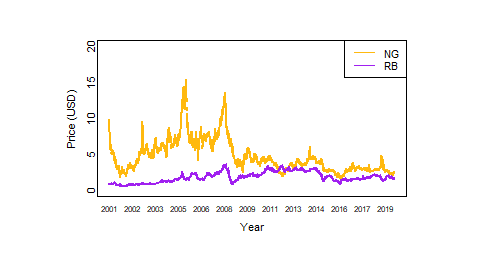}      
\caption{Prices (in USD) of the four assets (the scale on the $y$-axis varies between the two plots).}
      \label{fig:figure1a}
   \end{subfigure}
   \begin{subfigure}{0.99\textwidth}
      \centering
      \includegraphics[width=10cm, height=5cm]{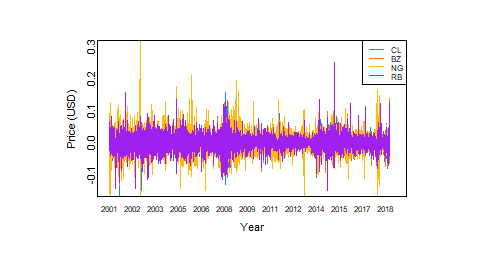}  
      \caption{Log-returns of the four assets.}
      \label{fig:figure1b}
   \end{subfigure}
   \caption{Assets during the period of analysis}
\end{figure}

The time series of BZ, CL, NG and RB asset prices and the correponding log-returns are shown in Figure \ref{fig:figure1a} and \ref{fig:figure1b}. We used daily data available at the website \url{https://www.backtestmarket.com}. All assets' dates are cross-matched to ensure the data series are synchronised with each other. Data refer to period from January 2001 through September 2019, with 4,608 data points. 
Table \ref{tab:descript_stats} provides descriptive statistics for the asset returns under analysis. The average log-returns are similar and the standard deviations are much larger than the corresponding means. Crude oil assets show a negative skewness, which can be interpreted as a tendency towards large decrease than large increase. Finally, the large values for the kurtosis statistics suggest distribution tails heavier than a normal distribution. 
Table \ref{tab:descript_tau} and Table \ref{tab:descript_rho} show the sample Kendall's $\tau$ and Spearman's $\rho$ for each pair of assets, which are measures of monotonic relationships among assets: as expected, the log-returns of the WTI and Brent are strongly correlated and RB is highly dependent on both WTI and Brent, being a derivative of crude oil. Finally NG seems to be only slightly correlated with the other assets. 

\begin{table}[h!]
  \begin{center}
    \begin{tabular}{l|c|c|c|c|c|c} 
      \textbf{Asset} & \textbf{Mean}  & \textbf{SD}  & \textbf{Max} & \textbf{Min} & \textbf{Skew.} & \textbf{Kurt.} \\
      \hline
      \textbf{BZ} & 2.068e-04 & $0.021$ & $0.136$ & $-0.134$ & $-0.080$ & $6.395$ \\
      \textbf{CL} & 1.668e-04 & $0.023$ & $0.157$ & $-0.165$ & $-0.131$ & $7.123$ \\
      \textbf{NG} & -2.598e-04 & $0.033$ & $0.324$ & $-0.325$ & $0.090$ & $10.464$ \\
      \textbf{RB} & 1.596e-04 & $0.025$ & $0.249$ & $-0.150$ & $0.116$ & $8.404$ \\
    \end{tabular}
    \caption{Descriptive statistics of the four asset returns.}
    \label{tab:descript_stats}
  \end{center}
\end{table}

\begin{table}[!htb]
    \begin{minipage}{.50\linewidth}
      \centering
    \begin{tabular}{l|c|c|c} 
      \textbf{$\tau$} &\textbf{BZ} & \textbf{CL} & \textbf{NG}   \\
      \hline
      \textbf{CL} & $0.753$ & {--} & {--} \\
      \hline
      \textbf{NG} & $0.164$ & $0.161$ & {--} \\
      \hline
      \textbf{RB} & $0.632$ & $0.596$ & $0.150$ \\
    \end{tabular}
    \caption{Sample Kendall's $\tau$ for each pair of assets.}
    \label{tab:descript_tau}
    \end{minipage}%
    \begin{minipage}{.50\linewidth}
      \centering
    \begin{tabular}{l|c|c|c} 
      \textbf{$\rho$} &\textbf{BZ} & \textbf{CL} & \textbf{NG}   \\
      \hline
      \textbf{CL} & $0.902$ & {--} & {--} \\
      \hline
      \textbf{NG} & $0.242$ & $0.237$ & {--} \\
      \hline
      \textbf{RB} & $0.811$ & $0.777$ & $0.221$ \\
    \end{tabular}
    \caption{Sample Spearman's $\rho$ for each pair of assets.}
    \label{tab:descript_rho}
    \end{minipage} 
\end{table}

\begin{figure}[h!]
  \centering
  \includegraphics[width=12cm,height=7cm]{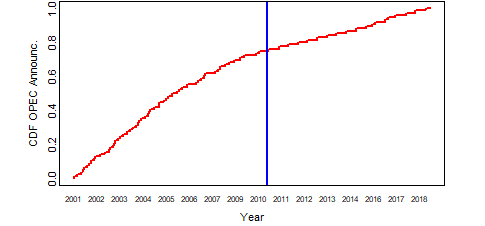}
  \caption{Empirical CDF of the OPEC announcements with a change in the production.}
  \label{fig:announc}
\end{figure}

OPEC announcements for the same period are available at \url{https://www.opec.org/opec\_web/en/press\_room/28.htm}. We have decided to give equal weight to each release of new information on the movement of prices (decision of increase or decrease of OPEC's production), in order to follow \cite{jiang2014testing} and show that energy market needs time to process new information; therefore a dummy variable is introduced, which is equal to one when an announcement is made, independently from the type of decision made, and zero for no announcement. 
Figure \ref{fig:announc} show the empirical cumulative distribution function for the binary time series: it can be noted that announcements occurred irregularly in the first period (Period 1), and regularly in the second one (Period 2); the blue line indicates the separation among the two periods. The changepoint has been inferred through a comparison in the means of the number of announcements per quarter: a t-test has been performed to compare the mean number of announcements in two consecutive quarters and the only significant difference correponded to the 37th quarters, which represents the first quarter of 2010. Therefore, we considered the changepoint in January 2010. 
 
\subsection{Marginal models}

It is well known that financial markets often exhibit time-varying volatility and this feature is well characterised by generalised autoregressive conditional heteroscedasticity (GARCH) models. Given a time series $\{Y_t\}_{t=0}^T$, its realisation follows a GARCH$(p,q)$ model if
\begin{align*}
    y_t &= \sigma_t \varepsilon_t \\
    \sigma_t &= \sqrt{\alpha_0 + \sum_{i=1}^p \alpha_i y_{t-1}^2 + \sum_{j=1}^q \beta_i \sigma_{t-j}^2},
\end{align*}
where $\varepsilon_t$ is a white noise, $p,q \geq 1$, and $\boldsymbol{\alpha}=(\alpha_1, \ldots, \alpha_p)$ and $\boldsymbol{\beta}=(\beta_1, \ldots, \beta_q)$ are parameters from an autoregressive (AR) and a moving average (MA) process respectively; see \cite{terasvirta2009introduction} for a review. 

There is evidence that crude oil price returns show heavy tails \citep{mohammadi2010international}, therefore we will consider a version of the GARCH$(p,q)$ model with Student-$t$ innovations, so that $y_t = \sigma_t \varepsilon_t$ where $\varepsilon_t \sim t(\nu)$, i.e. the error terms follows a $t$-distribution with $\nu$ degrees of freedom. It is possible to show that this model can be expressed via data augmentation \citep{geweke1993bayesian}:
\begin{align*}
    y_t &= \varepsilon_t \sqrt{\frac{\nu -2}{\nu} \omega_t \sigma_t^2} \\
    \sigma_t &= \sqrt{\alpha_0 + \sum_{i=1}^p \alpha_i y_{t-1}^2 + \sum_{j=1}^q \beta_i \sigma_{t-j}^2} \\
    \varepsilon_t &\sim \mathcal{N}(0,1) \\
    \omega_t &\sim IG\left( \frac{\nu}{2}, \frac{\nu}{2} \right)
\end{align*}
where $IG(a,b)$ stands for the inverse gamma distribution with shape parameter $a$ and scale parameter $b$. 

Considering $t$-distributed innovations allows to fit leptokurtic data, since the kurtosis is associated to the number of degrees of freedom. However, the distribution still cannot deal with skewness, which may characterize financial data. Therefore, it is possible to consider $\varepsilon_t$ from a skew-normal or a skew-$t$ distribution \citep{salisu2016modelling}, by introducing a skewness parameter $\lambda >0$.

We will perform model selection to identify which GARCH model is the most suitable for each marginal distribution and to properly select the values of $p$ and $q$. The main advantage to introduce a copula approach is that it allows to flexibly and realistically model each univariate time series separately and then model the dependence structure as a second step. In particular, GARCH models are known to suffer of the curse of dimensionality \citep{caporin2014robust}, however we overcome this problem by modelling the time series relative to each asset separately and then relate them through a copula model. 

There is evidence that misspecified marginal distributions can affect copula estimation, by increasing bias and mean-squared error \citep{azam2014effects,fantazzini2009effects}, however this effect is particularly evident for small sample sizes, and it decreases as the length of the time series increases, as it is also empirically shown in \cite{grazian2017approximate}. In order to check robustness with respect to the selected marginals, we will also implement copula estimation by using nonparametric estimation of the marginals. 

\subsection{Copula models}

Multivariate analysis can be affected by the assumptions implied in multivariate models. Multivariate distributions, such as the multivariate Gaussian distribution or the multivariate Student-$t$ distribution, usually assume that all the marginal distributions have the same form; moreover, the dependence structure is often assumed to be linear, while in many settings there could be asymmetries, in particular in the tails of the multivariate distributions, or the relationship among processes can be more generally monotonic and not necessarily linear \citep{embrechts2002correlation}.  

Copula models are representations of multivariate distribution which can take into account these generalisations. Given a random variable $\mathbf{Y}=(Y_1, \dots, Y_d)$ with $d$-variate cumulative distribution function (CDF) $\mathbf{F}$, Sklar's theorem \citep{sklar1959fonctions} proves that there exists a $d$-variate function $C:[0,1]^d \xrightarrow{} [0,1]$ such that
\begin{equation*}
    \mathbf{F}(y_1,\ldots, y_d;\phi_1, \ldots, \phi_d, \psi) = C(F_1(y_1;\phi_1), \ldots, F_d(y_d;\phi_d);\psi)
\end{equation*}
where $F_j$ is the marginal CDF of $Y_j$, indexed by marginal parameter $\phi_j$ and $\psi$ is the parameter characterising the copula function. If $\mathbf{F}$ is continuous, it can be shown that this representation is also unique and, if $\mathbf{F}$ admits a density function $\mathbf{f}$
\begin{footnotesize}
\begin{equation}
    \mathbf{f}(y_1, \dots, y_d; \phi_1, \ldots, \phi_d, \psi) = c(F_1(y_1;\phi_1), \ldots, F_d(y_d;\phi_d);\psi) \cdot f_1(y_1;\phi_1) \cdot \ldots \cdot f_d(y_d;\phi_d)
    \label{eq:cop_density}
\end{equation}
\end{footnotesize}
\noindent where $c$ is the derivative of $C$; Equation \eqref{eq:cop_density} shows that the copula function captures the full dependence structure of the multivariate joint distribution. 

There are several copula functions available in the literature. Let define $u_j=F_j(y_j;\phi_j)$ for $j=1, \ldots, d$. The Gaussian copula is defined as 
\begin{equation*}
    C_G(u_1, \ldots, u_d) = \boldsymbol{\Phi}(\Phi^{-1}(u_1),\ldots,\Phi^{-1}(u_d);R)
\end{equation*}
where $\boldsymbol{\Phi}$ is the CDF of a multivariate standard normal variable, $\Phi^{-1}$ is the inverse of a univariate standard normal CDF, and $\psi = R$ is a correlation matrix $R \in [-1,1]^{d \times d}$. The Gaussian copula has no tail dependence and it can be considered a generalisation of the multivariate normal distribution, with marginal distribution that can be not-normal. 

Similarly, the Student-$t$ copula is defined as 
\begin{equation*}
    C_{t}(u_1, \ldots, u_d) = \mathbf{T}(T^{-1}(u_1;\nu),\ldots, T^{-1}(u_d;\nu); \nu, R)
\end{equation*}
where $\mathbf{T}$ is a multivariate Student-$t$ CDF, $T^{-1}$ is the inverse of a univariate Student-$t$ CDF, and $\psi=(\nu,R)$ where $\nu$ are the degrees of freedom and $R$ is a correlation matrix, $R \in [-1,1]^{d \times d}$. Differently from the Gaussian copula, it allows symmetric non-zero tail dependence \citep{schloegl2005note}. As for the marginal distributions, it is possible to extend the Student-$t$ copula to allow for asymmetry, to obtain the skew-$t$ copula:
\begin{equation*}
    C_{skew T}(u_1, \ldots, u_d) = skewT(skewT^{-1}(u_1; \delta_1,\nu ),\ldots, skewT^{-1}(u_d;\delta_d, \nu);\nu, R, \boldsymbol{\delta}).
\end{equation*}
where $\psi=(\nu,R, \boldsymbol{\delta})$, $\nu$ being the number of degrees of freedom, $R$ being a correlation matrix, and $\boldsymbol{\delta} = (\delta_1, \ldots, \delta_d)$ being a vector of skewness parameters. The multivariate skew-$t$ distribution is defined as in \cite{azzalini2003distributions}.

Copula functions showing asymmetric tail dependence are the Clayton and the Gumbel copulas. The Clayton copula is defined as
\begin{equation*}
    C(u_1, \ldots, u_d)=\left[\sum_{j=1}^d u_j^{-\theta}-d+1\right]^{-\frac{1}{\theta}} 
\end{equation*}
where $\psi=\theta$ with $\theta \in [-1, \infty)\setminus \{0\}$; the Clayton copula shows positive dependence in the lower tail and no dependence in the upper tail. On the contrary, the Gumbel copula is defined as
\begin{equation*}
    C(u_1, \ldots, u_d)=\exp\left[-\left( \sum_{j=1}^d (-\log u_j)^{\theta}\right)^{\frac{1}{\theta}} \right] ,
\end{equation*}
with $\psi=\theta$ and $\theta \in [1,\infty)$; the Gumbel copula shows positive dependence in the upper tail and no dependence in the lower tail. 
Another example of Archimedean copula showing no tail dependence is the Frank copula, defined as
\begin{equation*}
    C(u_1, \ldots,u_d)= \frac{1}{\theta} \log \left[ 1 + \frac{\prod_{j=1}^d \exp(-\theta u_j)-1}{[\exp(-\theta)-1]^{d-1}}\right] 
\end{equation*}
with $\psi=\theta$ and $\theta \in \mathbb{R} \setminus \{0\}$.

Copula models can be made more complex, by introducing dependence on a covariate \citep{patton2006modelling}. The most important extension is considering time-varying copula parameters, i.e. the copula parameters is allowed to vary with time according to some model \citep{aielli2013dynamic,salvatierra2015dynamic}.

In many settings, some functionals of the dependence are of interest. In the following, we introduce bivariate functionals; extensions to the multivariate settings exist, however they may not be uniquely defined. The Spearman's $\rho$ between two variables $Y_1$ and $Y_2$ is the correlation coefficient among the transformed variables $U=F_1(y_1)$ and $V=F_2(y_2)$, assessing the relationship among the two variables as a monotonic function; it has a copula definition
\begin{equation*}
    \rho_S = 12\int_0^1 \int_0^1 C(u,v) du dv -3.
\end{equation*}
Similarly, the Kendall's $\tau$ is another measure of rank correlation defined as
\begin{equation*}
    \tau = 4\int_0^1 \int_0^1 C(u,v) dC(u,v) -1.
\end{equation*}
Both the Spearman's $\rho$ and the Kendall's $\tau$ are invaritant under monotone transformations and they range in $[-1,1]$, where the sign of the index indicates the direction of the association between variables.  

The Spearman's $\rho$ and the Kendall's $\tau$ are equal to zero if the variables are independent, however the fact that $\rho=\tau=0$ does not imply that the variables are independent. Moreover, these indices may be equal for different families of copula models: for example, the Kendall's $\tau$ are equal for all elliptical families \citep{fang2002meta} and the Spearman's $\rho$ is the same for the Student-$t$ and the Gaussian copula.

On the other hand, the mutual information measures the divergence between the joint distribution represented by a copula and the model of independence, represented by the product of the marginals and can assume different values for different families of copulas, depending on the strength of the dependence \citep{ebrahimi2014comparison}. The mutual information \cite{granger1994using} for jointly continuous random variables is defined as 
$$
I(u_1, u_2) = \int_{[0,1]} \int_{[0,1]} C(u_1, u_2) \log \left( \frac{C(u_1, u_2)}{u_1 \cdot u_2}\right) du_1 du_2
$$
and the mutual information index for absolutely continuous distributions is
\begin{equation}
\delta^2(C) = 1- e^{-2I(y_1,y_2)}. 
\label{eq:delta}
\end{equation} 
The mutual information index ranges in $[0,1]$. Alternatively, the normalised version of the Bhattacharya-Matusita-Hellinger measure of dependence is defined as
$$
S_{\rho} = \frac{1}{2} \int_{\mathcal{Y}} \int_{\mathcal{Y}} \left( c^{1/2}(F_1(y_1),F_2(y_2)) - (f_1(y_1) \cdot f_2(y_2))^{1/2}\right)^2 dy_1 dy_2
$$
where $c$ is the joint copula density and $f_j$ is the marginal density of the $j$-th variable, $j=1,2$. The index $S_{\rho}$ is equal to zero if and only if $C = F_1 \cdot F_2$. \cite{granger2004dependence} shows that $S_{\rho}$ has also a copula representation as
$$
S_{\rho} = \int_{[0,1]} \int_{[0,1]} [1-c^{1/2}(u_1,u_2)] du_1 dv_2.
$$
The Bhattacharya-Matusita-Hellinger measure of dependence is a nonparametric measure of dependence which is robust to nonlinearity of the observations, and measures the departure from independence. \cite{geenens2022hellinger} defined a Hellinger correlation coefficient as
\begin{equation}
\mathcal{H}^2 = 1 - \int_{[0,1]} \int_{[0,1]} \sqrt{c(u_1, u_2)} du_1 du_2
\label{eq:h2}
\end{equation} 
and found an $L_2$-consistent estimator, which will be used in this work. 

Finally, it is important to test if random vectors can be assumed mutually independent, in order to characterise the dependence structure as a multivariate distribution or, more easily, as a combination of bivariate distributions. \cite{bakirov2006multivariate} proposed a distance correlation based on functionals of the characteristic functions, which can deal with variables of different nature. We will use it to test mutual indpependence of random vectors. Consider $Y_1$ as a random vector in $\mathbb{R}^{d_1}$ and $Y_2$ a random vector in $\mathbb{R}^{d_2}$, where $d_1$ and $d_2$ are positive integers. Define the characteristic functions of $Y_1$ and $Y_2$ as $g_{Y_1}$ and $g_{Y_2}$ respectively, and the joint characteristic function as $g_{Y_1, Y_2}$. The measure 
\begin{align}
\mathcal{V}^2(Y_1, Y_2; w) &= \parallel g_{Y_1, Y_2} - g_{Y_1} \cdot g_{Y_2} \parallel^2_w \nonumber \\
&= \int_{\mathbb{R}^{d_1+d_2}} | g_{Y_1, Y_2} (y_1,y_2) - g_{Y_1}(y_1)g_{Y_2}(y_2)|^2 w(y_1,y_2) dy_1 dy_2  
\label{eq:vsq}
\end{align}
between the joint characteristic function and the product of the marginal characteristic functions (where $\parallel \cdot \parallel $ indicates the norm and $w$ is a weight function) can be used to test the hypothesis of independence
\begin{align*}
H_0 &: g_{Y_1, Y_2} = g_{Y_1} \cdot g_{Y_2} \\
H_1 &: g_{Y_1, Y_2} \neq g_{Y_1} \cdot g_{Y_2}.
\end{align*}
The measure $\mathcal{V}^2$ has the property to be zero if and only if $Y_1$ and $Y_2$ are independent. \cite{szekely2007measuring} studied the theoretical properties of this test statistic, in particular consistency. 

Finally, some studies show that, in particular in presence of volatile markets, tail dependence is useful to study the behaviour of extremes in finance \citep{ane2003dependence}. Differently from rank correlations, tail dependence indices describe the concordance in the tails of the bivariate distribution, i.e. the dependence of crude oil markets to move together up or down:
\begin{align}
    \lambda_U &= \lim_{v \rightarrow 1^{-}} \Pr(F_1(y_1) > v | F_2(y_2) > v) = \lim_{v \rightarrow 1^{-}} \frac{1-2v + C(v,v)}{1-v} \label{eq:lambdaU} \\
    \lambda_L &= \lim_{v \rightarrow 0^{+}} \Pr(F_1(y_1) \leq v | F_2(y_2) \leq v) = \lim_{v \rightarrow 0^{+}} \frac{C(v,v)}{u} \label{eq:lambdaL} 
\end{align}
provided that the limits exist. Equation \eqref{eq:lambdaU} represents the upper tail dependence index, while Equation \eqref{eq:lambdaL} represents the lower tail dependence index. 

\subsection{Methods for inference and testing.}

Inferential approaches to parameter estimation for copula models often rely on a two-step procedure \citep{patton2004ontheout,
cherubini2004copula,
kim2007comparison,
nikoloulopoulos2008dependence}: first parameters of the univariate marginals are estimated from separate univariate likelihoods and then the parameter of the copula model is estimating by pluggin-in the estimates obtained from the first step into the multivariate likelihood. Such approach allows for asymptotic normality and consistency \citep{francq2004maximum}.  \cite{joe2005asymptotic} studied the efficiency of the two-step procedure, showing good efficiency except for the cases of extreme dependence, which does not seem to be the case of the application under analysis, see Section \ref{sec:results}. More specifically, under regularity conditions, the relative efficiency of the two-step estimator with respect to the full likelihood estimator tends to one in case of independence of the marginals, and reaches its minimum in case of perfect dependence. 

The two-step procedure presents the flexibility to choose the model that best fit each time series. For a time series of length $T$, with observed random vectors $\mathbf{y}_1, \ldots, \mathbf{y}_d$, there are $d$ log-likelihood functions for the univariatiate marginals:
$$
\ell_j(\phi_j) = \sum_{t=1}^T \log f_j(y_{j,t}; \phi_j) \qquad j=1, \ldots, d 
$$
and the log-likelihood for the joint distribution is:
\begin{align*}
    \ell(\phi_1, \ldots, \phi_d, \psi; \mathbf{y}_1, \ldots, \mathbf{y}_d) &= \sum_{t=1}^T \left[ \log f_1(y_{1,t};\phi_1) + \ldots + \log f_d(y_{d,t};\phi_d)\right. \\
                & \left. \qquad + \log c(u_{1,t}, \ldots, u_{d,t}; \psi) \right]
\end{align*}
where $\phi_j$ represents the vector of parameters of the marginal distributions for $j=\{1,\ldots,d \}$ and $\psi$ represents the vector of parameter of the copula function; $u_{j,t} = F_j(y_{j,t};\phi_j)$ are the pseudo-observations. Performing a fully maximum-likelihood approach requires an increasing level of computational difficulty as the dimension increases, therefore the two-stage procedure can be implemented. The two-stage procedure is performed by first estimating the parameters of the marginal distributions to obtain estimates $(\hat{\phi}_1, \ldots, \hat{\phi}_d)$ and, therefore, $\hat{f}_{j}(y_{j,t};\hat{\phi}_{j})$; then define $\hat{u}_{j,t} = \hat{F}_j(y_{j,t};\hat{\phi}_j)$ and estimate:
\begin{equation*}
    \hat{\psi} = \arg \max_{\psi \in \Psi} \sum_{t=1}^T \log c\ \left(\hat{u}_{1,t}, \ldots, \hat{u}_{d,t}; \psi \right).
\end{equation*}
Under regularity conditions, $(\hat{\phi}_1, \ldots, \hat{\phi}_d, \hat{\psi})$ is the solution of 
$$
\left( \frac{\partial \ell_1}{\partial \phi_1^T} , \ldots, \frac{\partial \ell_d}{\partial \phi_d^T}, \frac{\partial \ell}{\partial \psi^T}\right) = \mathbf{0}^T
$$
which is, in general, different from the maximum likelihood estimators given by the solutions to
$$
\left( \frac{\partial \ell}{\partial \phi_1^T} , \ldots, \frac{\partial \ell}{\partial \phi_d^T}, \frac{\partial \ell}{\partial \psi^T}\right) = \mathbf{0}^T.
$$

Selecting the correct model under the marginals may not be an easy task. In order to reduce the impact of model assumptions, several models can be compared for both the marginal distributions and the joint distribution, and model selection measures used to select the best model; for example, BIC (Bayesian information criterion) is often applied in this setting. However, selection criteria may disagree and sometimes there is not enough information to select a model with a reasonable degree of confidence.  \cite{fantazzini2009effects} studied the effect of misspecification of the marginals on the copula estimation: in particular, \cite{fantazzini2009effects} found via simulation that, when observations are characterised by skewness and symmetric marginals are assumed together with a Gaussian copula, correlations are negatively biased; the bias seems to increase when the true copula is a Student-$t$ copula and marginals do not properly account for skewness. Moreover, bias is still evident when the true copula generating the data is not elliptical and marginals do not account for skewness, although the sign of the bias may differ. 

\cite{genest1995semiparametric} proposed a semiparametric approach, where nonparametric estimates are obtained for the marginal distributions, while a parametric model is assumed for the copula. Then the copula parameter $\psi$ is estimated as that value $\hat{\psi}$ that maximises the pseudo log-likelihood
$$
\ell(\psi; \mathbf{y}_1, \ldots, \mathbf{y}_d) = \sum_{t=1}^T \log \left[ c\left( \hat{F}_{1n}(y_{1,t}) , \ldots, \hat{F}_{dn}(y_{d,t}) \right)\right]
$$
where $\hat{F}_{jn}(y_{j,t})=\frac{1}{T+1} \sum_{t=1}^T \mathbb{I}[Y_{j,t} < y_{j,t}]$ is the marginal empirical CDF of the $j$-th variable. The denominator is taken to be equal to $(T+1)$ so that difficulties arising when the estimates are close to one are avoided. The semiparametric approach can be more flexible because it does not need to choose the marginal parametric families and is more robust to model misspecification. The semiparametric estimator of the copula parameter is not as efficient as the maximum likelihood estimator \citep{genest1995semiparametric}, however \cite{genest2002conditions} studied the conditions for which it is asymptotically efficient. \cite{kim2007comparison} showed through simulations that, when marginals are misspecified, the mean squared error obtained by applying the two-step parametric procedures (and also the fully maximum likelihood approach) can be large, with respect to the semiparametric procedure, which is not computationally more expensive. 

In this work, we compare the results obtained by using either a fully parametric and a semiparametric approach. 


\section{Results}

\label{sec:results}

\subsection{Marginal models}
\label{sub:marginals}

\begin{figure}[h]
  \centering
  \includegraphics[scale=0.6]{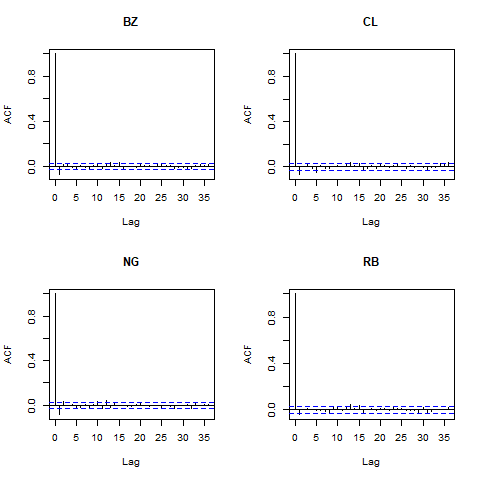}
  \caption{Autocorrelation plots for the four log-returns time series.}
  \label{fig:acf}
\end{figure}

First, log-returns for all time series have been tested for heteroscedasticity by using the Engle’s ARCH test \citep{engle1982autoregressive}; the test rejects the null hypothesis of no heteroscedasticity at lag $m$ for a significance level of 0.05. Similarly, we have applied the Ljung-Box Q-test \citep{ljung1978measure} for testing for autocorrelation and the null hypothesis of no autocorrelation at lag $m$ is rejected at a significance level of 0.05. Both tests have been applied for every lag $m \in \{1,2,\ldots, 200\}$; the Engle's ARCH test confirms persistence of heteroskedasticity for every time series at large lag (p-value of order $10^{-16}$); on the other hand, the Ljung-Box test for the hypothesis of independence in each times series shows no autocorrelation after lag $m=1$ for BZ (p-value of order $10^{-6}$), CL (p-value of order $10^{-5}$), NG (p-value of order $10^{-8}$) and after lag $m=13$ for RB (p-value of order $10^{-6}$).  Figure \ref{fig:acf} shows the autocorrelation plots for each series, suggesting that autocorrelation is not significantly different from zero after lag $m=1$; moreover, for RB autocorrelation, even if tested present, seems to be low. 

\begin{figure}[h]
  \centering
  \includegraphics[scale=0.6]{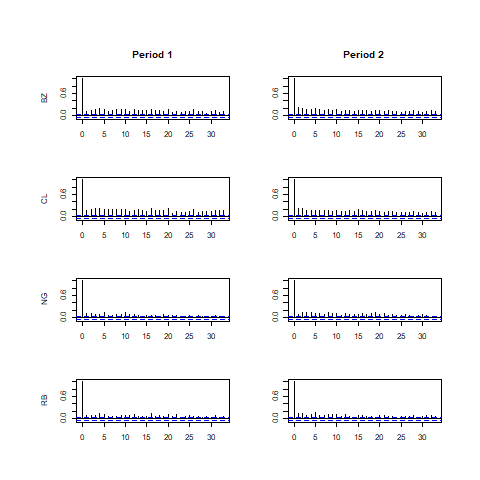}
  \caption{Autocorrelation plots for the absolute value of the four time series.}
  \label{fig:acf_abs}
\end{figure}

\begin{figure}[h]
  \centering
  \includegraphics[scale=0.6]{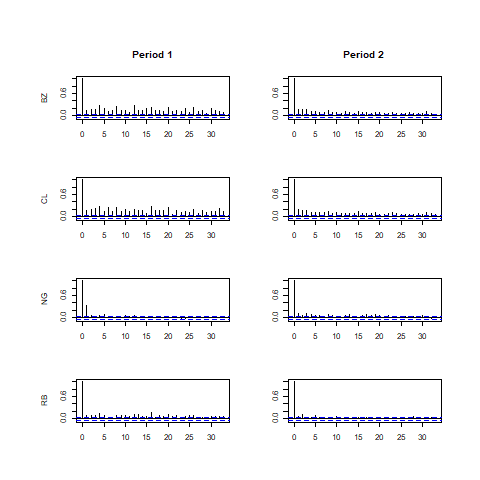}
  \caption{Autocorrelation plots for the square value of the four time series.}
  \label{fig:acf_sq}
\end{figure}

Figure \ref{fig:acf_abs} and \ref{fig:acf_sq} show the autocorrelation plots of the absolute value and the squared value of the time series of the four log-returns in the two time windows to investigate the persistence of volatility: they both agree that autocorrelation seems to remain at large lag, in particular for BZ and CL. 

\begin{figure}[h!]
  \centering
  \includegraphics[scale=0.5]{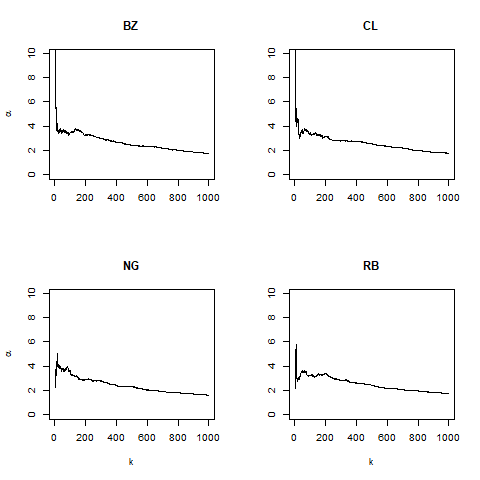}
  \caption{Plots of the Hill's estimators computed for different values of the upper-order statistics $k$.}
  \label{fig:hill}
\end{figure}

\begin{figure}[h!]
  \centering
  \includegraphics[scale=0.5]{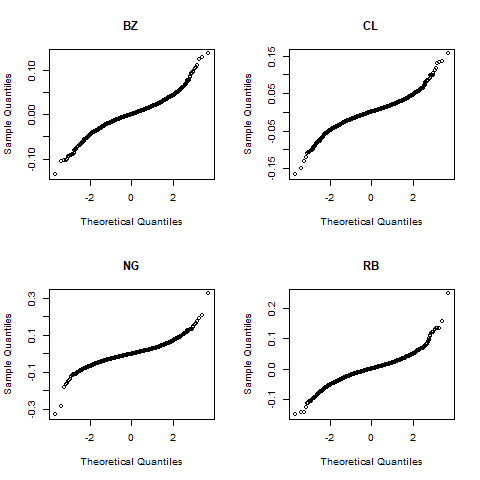}
  \caption{Qq-plots of the four time series.}
  \label{fig:qqplot}
\end{figure}

In order to test for heavy tails, the Hill's estimator has been computed for each time series:
$$
\hat{\alpha} = \left[ \frac{1}{k} \sum_{i=1}^{k-1} \log(y_{T-i} - \log(y_{T-k}))\right]^{-1}
$$
where $k$ is the number of upper-order statistics used in the estimation. Figure \ref{fig:hill} shows the obtained plots for values of $k \in \{1,\ldots, 1000\}$, showing that the estimates of the Hill's index are always larger than zero, for all the considered values of $k$. Moreover, Figure \ref{fig:qqplot} shows the qq-plots for each time series, from which it is evident that the tails of each time series do not allow for a Gaussian assumption.  

We have fitted GARCH($p,q$) models with Gaussian and Student-$t$ innovation, and allowing or not for skewness, with parameters $p$ and $q$ ranging from zero to ten. The best model for each time series has been selected via Bayesian information criterion (BIC) for all models, see Table \ref{tab:aic_bic_gar}; for reasons of space, only values of $p \in \{1,2\}$ and $q\in \{0,1,2\}$ are shown, since the values of BIC for all the other models where larger. In all cases, models with $p=1$ and $q=1$ and Student-$t$ innovations are preferred, which means that extreme returns are more likely than in the Gaussian case. Moreover, the time series of BZ and CL in the second time window and of NG in both time windows seem to show asymmetry, with a preference for skew-$t$ models. It is also possible to see from Table \ref{tab:aic_bic_gar} that the BIC values are similar for all models, which means that data are not strongly supporting the selected models against the others. For this reason, in Section \ref{sub:copula} a comparison between parametrically estimated marginal and nonparametrically estimated marginals and their effect on the estimation of the copula parameters and the dependence measures is provided. 

Table \ref{tab:margin_estim} shows the maximum likelihood estimates of the parameters of marginal models selected as in Table \ref{tab:aic_bic_gar}, in the period before and after the change in the regularity of the OPEC announcements. The possibility to separately model the marginal distributions, which is an essential part of the copula approach, allows to use the models that show the better fit to the marginal data: for example, it is possible to use models with asymmetry for BZ and CL in Period 2 and NG in both Periods, and models without asymmetry for the other time series. Moreover, it is possible to model the log-returns with flexible models, as the GARCH-skew-$t$ model, without incurring in the curse of dimensionality previously described, since we only model the one-dimensional distributions. 

The comparison among periods shows that there are slight differences among the parameters of each univariate distribution; the parameters which shows the largest change in the estimation is the number of degrees of freedom $\nu$. It is interesting to notice that the number of degrees of freedom for BZ, CL and RB is decreasing between the two time windows, while it is increasing for NG. The number of degrees of freedom in a Student-$t$ or skew-$t$ distribution is associated with the kurtosis of the distribution: lower values of $\nu$ are associated with leptokurtic distributions (heavy tails) and larger values of $\nu$ are associated with vanishing kurtosis. The results show that heavier tails are more suitable for model crude oil prices in the second time window, while NG shows lighter tails. With a copula approach, it is possible to separately estimate the degrees of freedom of each time series and it is then possible to flexibly model the tails of each univariate distribution. 

\pagebreak

\begin{tiny}
\begin{center}
  \begin{longtable}{c|c| c c c |c|c c c }
    & \textbf{Period 1} & q $= 0$ & q $= 1$ & q $= 2$ & \textbf{Period 2} & q $= 0$ & q $= 1$ & q $= 2$ \\
    \hline
    \textbf{BZ} & \multicolumn{7}{l}{Gaussian} \\
    & p $= 1$ & -4.713 & -4.844 & -4.841 & p $= 1$ & -5.093 & -5.252 & -5.248 \\
    & p $= 2$ & -4.744 & -4.840 & -4.837 & p $=2$ & -5.126 & -5.248 & -5.245 \\
    & \multicolumn{7}{l}{skew-Gaussian} \\
    & p $= 1$ & -4.711 & -4.842 & -4.839 & p $= 1$ & -5.091 & -5.257 & -5.254 \\
    & p $= 2$ & -4.742 & -4.838 & -4.835 & p $=2$ & -5.124 & -5.253 & -5.250 \\
    & \multicolumn{7}{l}{Student-$t$} \\
    & p $= 1$ & -4.774 & \textbf{-4.852} & -4.848 & p $= 1$ & -5.229 & -5.329 & -5.327 \\
    & p $= 2$ & -4.784 & -4.848 & -4.845 & p $=2$ & -5.250 & -5.326 & -5.323 \\
    & \multicolumn{7}{l}{skew-$t$} \\
    & p $= 1$ & -4.771 & -4.849 & -4.845 & p $= 1$ & -5.228 & \textbf{-5.330} & -5.327 \\
    & p $= 2$ & -4.781 & -4.845 & -4.842 & p $=2$ & -5.249 & -5.327 & -5.324 \\
      
    \hline
    \textbf{CL} & \multicolumn{7}{l}{Gaussian} \\
    & p $= 1$ & -4.547 & -4.692 & -4.691 & p $= 1$ & -4.947 & -5.094 & -5.090 \\
    & p $= 2$ & -4.585 & -4.689 & -4.687 & p $=2$ & -4.992 & -5.090 & -5.088 \\
    & \multicolumn{7}{l}{skew-Gaussian} \\
    & p $= 1$ & -4.546 & -4.693 & -4.691 & p $= 1$ & -4.946 & -5.102 & -5.099 \\
    & p $= 2$ & -4.587 & -4.690 & -4.688 & p $=2$ & -4.992 & -5.099 & -5.096 \\
    & \multicolumn{7}{l}{Student-$t$} \\
    & p $= 1$ & -4.624 & \textbf{-4.712} & -4.708 & p $= 1$ & -5.044 & -5.149 & -5.145 \\
    & p $= 2$ & -4.636 & -4.709 & -4.706 & p $=2$ & -5.070 & -5.145 & -5.143 \\
    & \multicolumn{7}{l}{skew-$t$} \\
    & p $= 1$ & -4.621 & -4.709 & -4.706 & p $= 1$ & -5.047 & \textbf{-5.155} & -5.151 \\
    & p $= 2$ & -4.633 & -4.707 & -4.703 & p $=2$ & -5.073 & -5.151 & -5.148 \\
      
    \hline
    \textbf{NG} & \multicolumn{7}{l}{Gaussian} \\
    & p $= 1$ & -3.738 & -3.806 & -3.804 & p $= 1$ & -4.370 & -4.509 & -4.506 \\
    & p $= 2$ & -3.747 & -3.803 & -3.800 & p $=2$ & -4.377 & -4.508 & -4.505 \\
    & \multicolumn{7}{l}{skew-Gaussian} \\
    & p $= 1$ & -3.744 & -3.813 & -3.810 & p $= 1$ & -4.367 & -4.507 & -4.503 \\
    & p $= 2$ & -3.754 & -3.810 & -4.808 & p $=2$ & -4.375 & -4.506 & -4.502 \\
    & \multicolumn{7}{l}{Student-$t$} \\
    & p $= 1$ & -3.824 & -3.872 & -3.869 & p $= 1$ & -4.461 & -4.529 & -4.526 \\
    & p $= 2$ & -3.833 & -3.871 & -3.868 & p $=2$ & -4.473 & -4.531 & -4.528 \\
    & \multicolumn{7}{l}{skew-$t$} \\
    & p $= 1$ & -3.822 & \textbf{-3.873} & -3.870 & p $= 1$ & -4.458 & \textbf{-4.530} & -4.522 \\
    & p $= 2$ & -3.832 & -3.872 & -3.868 & p $=2$ & -4.470 & -4.526 & -4.525 \\
      
    \hline
    \textbf{RB} & \multicolumn{7}{l}{Gaussian} \\
    & p $= 1$ & -4.289 & -4.348 & -4.345 & p $= 1$ & -4.911 & -5.006 & -5.002 \\
    & p $= 2$ & -4.295 & -4.345 & -4.342 & p $=2$ & -4.942 & -5.003 & -5.006 \\
    & \multicolumn{7}{l}{skew-Gaussian} \\
    & p $= 1$ & -4.290 & -4.347 & -4.344 & p $= 1$ & -4.908 & -5.003 & -4.999 \\
    & p $= 2$ & -4.295 & -4.344 & -4.340 & p $=2$ & -4.938 & -5.000 & -5.003 \\
    & \multicolumn{7}{l}{Student-$t$} \\
    & p $= 1$ & -4.347 & \textbf{-4.386} & -4.383 & p $= 1$ & -5.067 & \textbf{-5.115} & -5.112 \\
    & p $= 2$ & -4.350 & -4.383 & -4.379 & p $=2$ & -5.074 & -5.112 & -5.108 \\
    & \multicolumn{7}{l}{skew-$t$} \\
    & p $= 1$ & -4.346 & -4.385 & -4.381 & p $= 1$ & -5.067 & -5.114 & -5.111 \\
    & p $= 2$ & -4.350 & -4.382 & -4.378 & p $=2$ & -5.072 & -5.111 & -5.107 \\
    \caption{BIC values for combinations of $p$ and $q$ in GARCH models of log-returns for Gaussian, skew-normal, Student-$t$, and skew-$t$ innovations. The selected model (the model with the lowest BIC) is given in bold. Models with $p>2$ and $q>2$ are not provided because they are associated with larger BIC values.}
    \label{tab:aic_bic_gar}
  \end{longtable}
\end{center}
\end{tiny}

\begin{table}[]
\centering
\begin{tabular}{c|cc|c|cc}
\textbf{BZ} & \textbf{Period 1}    & \textbf{Period 2}    & \textbf{CL} & \textbf{Period 1}    & \textbf{Period 2}    \\ \hline
$\alpha_0$  & 7.908e-06            & 2.011e-06            & $\alpha_0$  & 8.529e-06            & 2.614e-06            \\
            & \textit{(2.937e-06)} & \textit{(3.114e-04)} & \textit{}   & \textit{(2.911e-06)} & \textit{(1.163e-06)} \\
$\alpha_1$  & 0.049                & 0.061                & $\alpha_1$  & 0.047                & 0.060                \\
            & \textit{(0.010)}     & \textit{(0.012)}     & \textit{}   & \textit{(0.009)}     & \textit{(0.010)}     \\
$\beta_1$   & 0.935                & 0.937                & $\beta_1$   & 0.938                & 0.937                \\
            & \textit{(0.013)}     & \textit{(0.011)}     & \textit{}   & \textit{(0.001)}     & \textit{(0.010)}     \\
$\nu$       & 10.000               & 5.427                & $\nu$       & 10.000               & 6.520                \\
            & (1.590)              & (0.599)              &             & (1.599)              & \textit{(0.925)}     \\
$\lambda$   & -                 & 0.921                & $\lambda$   & -                  & 0.874                \\
            & -                  & \textit{(0.026)}     &             & -                  & \textit{(0.025)}     \\ \hline
\textbf{NG} & \textbf{Period 1}    & \textbf{Period 2}    & \textbf{RB} & \textbf{Period 1}    & \textbf{Period 2}    \\ \hline
$\alpha_0$  & 3.451e-05            & 1.685e-05            & $\alpha_0$  & 1.111e-05            & 2.461e-06            \\
            & \textit{(1.034e-05)} & \textit{(4.633e-06)} & \textit{}   & \textit{(4.835e-06)} & \textit{(1.199e-06)} \\
$\alpha_1$  & 0.071                & 0.076                & $\alpha_1$  & 0.033                & 0.027                \\
            & \textit{(0.012)}     & \textit{(0.011)}     & \textit{}   & \textit{(0.008)}     & \textit{(0.007)}     \\
$\beta_1$   & 0.906                & 0.901                & $\beta_1$   & 0.953                & 0.968                \\
            & \textit{(0.014)}     & \textit{(0.013)}     & \textit{}   & \textit{(0.012)}     & \textit{(0.009)}     \\
$\nu$       & 6.074                & 9.472                & $\nu$       & 7.242                & 4.841                \\
            & \textit{(0.740)}     & \textit{(1.647)}     & \textit{}   & \textit{(1.015)}     & \textit{0.459}       \\
$\lambda$   & 1.095                & 1.024                & $\lambda$   & -                  & -                  \\
            & \textit{(0.032)}     & \textit{(0.003)}     &             & -        & -   \\              
\end{tabular}
\caption{Parameter maximum likelihood estimates for the models selected as optimal models with BIC as in Table \ref{tab:aic_bic_gar} (standard errors in brackets): $\alpha_0$ is the location of the volatility model, $\alpha_1$ and $\beta_1$ are the coefficient of the AR part and the MA part respectively, $\nu$ is the number of degrees of freedom for Student-$t$ innovations, and $\lambda$ is the parameter of asymmetry for skewed innovations.}
\label{tab:margin_estim}
\end{table}

\subsection{Copula model}
\label{sub:copula}

We now analyse the change in the dependence structure among the four assets considered in this work before and after the change in the regularity of the OPEC announcements. First, we investigate the structure of mutual independence of the four assets with the test proposed by \cite{bakirov2006multivariate} and \cite{szekely2007measuring} and defined in Equation \eqref{eq:vsq}. The test statistic $\mathcal{V}^2$ is 0.501 when testing the independence between the bivariate distribution of BZ and CL and the bivariate distribution of NG and RB (p-value: 0.001), 0.523 for BZ-NG versus CL-RB (p-value: 0.001), and 0.530 for CL-NG versus BZ-RB (p-value: 0.001), suggesting that also the bivariate distributions are dependent. Similarly, the test statistic is 0.524 (p-value: 0.001) for the trivariate distribution of (BZ,CL,NG) against the univariate distribution of RB, 0.577 (p-value: 0.001) for (CL,NG,RB) against BZ, 0.092 (p-value: 0.001) for (BZ,CL,RB) against NG, and 0.572 (p-value: 0.001) for (BZ,NG,RB) against CL, again suggesting dependence among the trivariate distributions and the univariate distributions. 

We have fit six different copula models (Gaussian, Student-$t$, skew-$t$, Clayton, Frank and Gumbel copula). Table \ref{tab:bic4} shows the BIC values for the considered copula function for the four-dimensional distribution of the four log-returns time series, obtained with nonparametric estimation of the marginal distributions. In both periods, the selected copula models is a $t$-copula and the estimates for the correlation matrix $R$ is available in Table \ref{tab:rhomat}: it is possible to notice a reduction of the correlation between Period 1 and Period 2, in particular for those pairs of assets involving NG (for which the correlation is already low in Period 1). The number of degrees of freedom passes from 5.427 estimated in Period 1, to 5.139 estimated in Period 2. 

\begin{table}[h]
\centering
\begin{tabular}{cc|cc}
\textbf{Period 1} & \textbf{BIC}                        & \textbf{Period 2} & \textbf{BIC}                        \\ \hline
\textit{Gaussian} & -3800.966                           & \textit{Gaussian} & -2611.677                           \\
\textit{$t$}      & \textbf{-4571.084} & \textit{$t$}      & \textbf{-7391.888} \\
\textit{skew-$t$} & -2918.942                           & \textit{skew-$t$} & -4345.774                           \\
\textit{Clayton}  & -3132.900                           & \textit{Clayton}  & -2469.057                           \\
\textit{Frank}    & -3450.927                           & \textit{Frank}    & -2306.257                           \\
\textit{Gumbel}   & -3439.927                           & \textit{Gumbel}   & -2209.688                          
\end{tabular}
\caption{BIC values for several copula functions for the four-dimensional distribution of the four assets. The bold values are relative to the smallest value of BIC.}
\label{tab:bic4}
\end{table}

\begin{table}[h]
\centering
\begin{tabular}{l|llll|l|llll}
 & \multicolumn{4}{l|}{\textbf{Period 1}}        & & \multicolumn{4}{l}{\textbf{Period 2}}    \\
                  & BZ    & CL    & NG    & RB    &                   & BZ    & CL    & NG    & RB    \\ \hline
BZ                & 1.000 & 0.945 & 0.313 & 0.806 & BZ                & 1.000 & 0.922 & 0.121 & 0.871 \\
CL                &       & 1.000 & 0.312 & 0.808 & CL                &       & 1.000 & 0.125 & 0.811 \\
NG                &       &       & 1.000 & 0.290 & NG                &       &       & 1.000 & 0.110 \\
RB                &       &       &       & 1.000 & RB                &       &       &       & 1.000
\end{tabular}
\caption{Values of the $R$ correlation matrix of the $t$ copula selected in Table \ref{tab:bic4}.}
\label{tab:rhomat}
\end{table}

The definition of multivariate functionals of the dependence is problematic, since several definitions exist for the same functional \cite{schmid2007multivariate}. Therefore, we now investigate the bivariate copulas, in order to better characterise the dependence structure. 
 
We fit the same six copula models for each pair of assets, such that
\begin{equation*}
    \hat{\psi}_{jk} = \arg \max_{\psi_{jk} \in \Psi} \sum_{t=1}^T \log c\ \left(\hat{u}_{j,t},\hat{u}_{k,t}; \psi_{jk} \right).
\end{equation*}
for $j,k=\{BZ,CL,RB,NG\}$ and where $\hat{u}_{j,t}$ are the pseudo-observations obtained either by plugging it the parameter estimates of the marginal models or as empirical CDF estimates; the parameter $\psi$ varies according to the specific copula model: for the Gaussian copula, $\psi = \rho$, where $\rho$ is the correlation parameter between the two assets; for the Student-$t$ copula, $\psi=(\rho,\nu)$, where $\rho$ is the correlation coefficient between the two assets and $\nu$ is the number of degrees of freedom; for the skew-$t$ copula, $\psi=(\rho,\delta,\nu)$ where $\rho$ is the correlation parameter between the two assets, $\nu$ is the number of degrees of freedom and $\delta$ is the skewness parameter; for the Clayton, Gumbel and Frank copulas $\psi=\theta$ where $\theta$ is the parameter for each of the Archimedean copulas. 

For each copula, we have computed the BIC to select the best model (Table \ref{tab:np_copulafit_BIC} and \ref{tab:param_copulafit_BIC}). We implemented also dynamic-copula models, where the parameters of the copula are allowed to vary over time, using the \texttt{Dynamic Copula Toolbox 3.0} \citep{vogiatzoglou2021}. In every case, the non-dynamic version was preferred according to the BIC, therefore we have omitted those results. Table \ref{tab:np_copulafit_BIC} shows the BIC values when marginals are nonparametrically estimated, while Table \ref{tab:param_copulafit_BIC} shows the BIC values for copula models when marginals are parametrically estimated: it is possible to see that the same model is selected in all cases, which means that, in this particular example, the parametric estimation of the marginals seems to be robust to the selection of the copula model. 

\begin{center}
  \begin{longtable}{l| c c c c c c c}
    & \textbf{Gaussian} & \textbf{Student-$\textit{t}$} & \textbf{Skew-$\textit{t}$} & \textbf{Clayton} & \textbf{Frank} & \textbf{Gumbel} \\
    \hline
    \multicolumn{7}{l}{\textbf{Period 1}} \\
    \hline
    BZ-CL	&	-4117.66	&	\textbf{-5190.26}	& -2794.03	&	-3894.68	&	-4354.94	&	-4593.52		\\
    BZ-NG	&	-310.65	&	\textbf{-321.23}	& -189.90 	&	-120.56		&	-300.48		&	-298.49		\\
    BZ-RB	&	-2429.25	&	\textbf{-2619.31}	& -1110.50 	&	-1990.35		&	-2286.32		&	-2427.56		\\
    CL-NG	&	-290.38	&	\textbf{-302.15} 	& -188.33 	&	-167.23		&	-287.10		&	-269.34		\\
    CL-RB	&	-2364.18	&	\textbf{-2652.75}	& -1559.21 	&	-1976.45		&	-2316.80		&	-2434.86	\\
    NG-RB	&	-264.13	&	\textbf{-278.10}	& -207.81 	&	-51.67		&	-246.97		&	-243.21		\\
    \hline
    \multicolumn{7}{l}{\textbf{Period 2}} \\
    \hline
    BZ-CL	&	-4031.15	&	\textbf{-4391.16}	& -2454.40		 &	-3712.93			&	-3755.15		&	-4037.03	\\
    BZ-NG	&	-25.59	&	-27.75	          	& -10.60 			&	\textbf{-41.40}	&	-19.87		&	-15.94		\\
    BZ-RB	&	-2796.11	&	\textbf{-3097.66}	& -1738.28 		&	-2475.06			&	-2833.84		&	-2826.30		\\
    CL-NG	&	-25.26	&	-31.88	           	& -10.96 			&	\textbf{-50.26}	&	-20.15		&	-13.33		\\
    CL-RB	&	-2118.41	&	\textbf{-2284.69}	& -945.29 		&	-1873.64			&	-2039.89	&	-2077.77		\\
    NG-RB	&	-20.55	&	-25.47	            	&  -8.91 			&	\textbf{-31.56}	&	-13.07		&	-17.61		\\
    \caption{BIC competing copula models with nonparametric estimation of the marginals.}
    \label{tab:np_copulafit_BIC}
  \end{longtable}
\end{center}

\begin{center}
  \begin{longtable}{l| c c c c c c c}
    & \textbf{Gaussian} & \textbf{Student-$\textit{t}$} & \textbf{Skew-$\textit{t}$} & \textbf{Clayton} & \textbf{Frank} & \textbf{Gumbel} \\
    \hline
    \multicolumn{7}{l}{\textbf{Period 1}} \\
    \hline
    BZ-CL	&	-4417.92	&	\textbf{-5561.25}		& -5558.12 	&	-4261.75		&	-4897.02		&	-4922.33		\\
    BZ-NG	&	-293.28	&	\textbf{-300.42}	    	& -294.58 	&	-199.84		&	-294.88		&	-282.30		\\
    BZ-RB	&	-2642.27	&	\textbf{-2841.43}		& -2835.09 	&	-2180.52		&	-2668.69		&	-2606.24		\\
    CL-NG	&	-268.89	&	\textbf{-278.71} 		& -271.11 	&	-196.40		&	-273.92		&	-251.74		\\
    CL-RB	&	-2551.66	&	\textbf{-2872.38}		& -2865.59 	&	-2171.51		&	-2668.63		&	-2610.65		\\
    NG-RB	&	-251.17	&	\textbf{-258.80} 	 	& -251.12		&	-191.64		&	-248.26		&	-229.97		\\
    \hline
    \multicolumn{7}{l}{\textbf{Period 2}} \\
    \hline
    BZ-CL	&	-3606.11	&	\textbf{-3923.62}		& -2924.94	&	-3283.16			&	-3241.28		&	-3599.84	\\
    BZ-NG	&	-24.10	&	-21.01	  		          & -26.27		&	\textbf{-41.90}	&	-18.43		&	-8.18		\\
    BZ-RB	&	-2471.42	&	\textbf{-2779.69}		& -2774.71 	&	-2193.09			&	-2391.46		&	-2522.41		\\
    CL-NG	&	-27.11	&	-27.50	 	 	          & -40.62 	&	\textbf{-54.05}	&	-22.18		&	-7.52		\\
    CL-RB	&	-1841.31	&	\textbf{-1996.09}		& -1994.79 	&	-1628.17			&	-1690.92		&	-1801.72		\\
    NG-RB	&	-18.17	&	-13.61	  	 	         &  -13.31 		&	\textbf{-29.26}	&	-11.46		&	-7.71		\\
    \caption{BIC competing copula models with parametric estimation of the marginals, with the best model chosen as in Table \ref{tab:np_copulafit_BIC}.}
    \label{tab:param_copulafit_BIC}
  \end{longtable}
\end{center}

The Student-$t$ copula is chosen as the best model for most of 
the pairs of assets in Period 1. In the second period, the Student-$t$ is the best model for asset pairs not involving NG, and the Clayton copula is the best for asset pairs involving NG. The latter model suggests that in the second period, pair of assets involving NG show no upper tail dependence. However, these are cases where the dependence is quite weak, in general. 

Table \ref{tab:copulafit_parameters} shows the point estimates and the relative standard errors of the model selected as optimal by BIC for each pair of assets. Gaussian, Gumbel, Frank and skew copulas are not shown in Table \ref{tab:copulafit_parameters} because they were never selected. Table \ref{tab:copulafit_parameters} provides estimates obtained with nonparametric estimation of the marginals; results obtained with parametric estimation of the marginals were very similar and, for this reason, omitted. 

Table \ref{tab:copulafit_parameters} shows the estimated parameters of given copulas for each pair of assets which are calculated using a maximum likelihood estimator. The parameter $\rho$ represents the linear correlation parameter in the elliptical copulas, and $\nu$ represents the number of degrees of freedom for the Student-$t$ copula. For the bivariate Clayton copula, $\theta$ is the copula parameter. 

\begin{table}[]
\begin{tabular}{c|ccc|ccc}
\multicolumn{1}{l|}{} & \multicolumn{1}{l}{\textbf{Period 1}} & \multicolumn{1}{l}{\textbf{}} & \multicolumn{1}{l|}{\textbf{}} & \multicolumn{1}{l}{\textbf{Period 2}} & \multicolumn{1}{l}{} & \multicolumn{1}{l}{} \\ \cline{2-7}
\multicolumn{1}{l|}{} & \multicolumn{1}{l}{Student-$t$} & \multicolumn{1}{l}{\textbf{}} & \multicolumn{1}{l|}{Clayton} & \multicolumn{1}{l}{Student-$t$} & \multicolumn{1}{l}{\textbf{}} & \multicolumn{1}{l}{Clayton} \\ 
                      & $\rho$                                & $\nu$                         & $\theta$                       & $\rho$                                & $\nu$                & $\theta$             \\ \hline
\textbf{BZ-CL}        & 0.947                                 & 1.976                         & -                              & 0.918                                 & 2.393                & -                    \\
\textbf{}             & \textit{(0.017)}                               & \textit{(3.003)}                       &                                & \textit{(0.004)}                               & \textit{(0.241)}              &                      \\ \hline
\textbf{BZ-NG}        & 0.327                                 & 13.515                        & -                              & -                                     & -                    & 0.202                \\
\textbf{}             & \textit{(0.017)}                               & \textit{(3.546)}                       &                                &                                       &                      & \textit{(0.032)}              \\ \hline
\textbf{BZ-RB}        & 0.839                                 & 4.500                         & -                              & 0.875                                 & 3.721                & -                    \\
\textbf{}             & \textit{(0.008)}                               & \textit{(0.486)}                       &                                & \textit{(0.005)}                               & \textit{(0.387)}              &                      \\ \hline
\textbf{CL-NG}        & 0.319                                 & 14.136                        & -                              & -                                     & -                    & 0.138                \\
\textbf{}             & \textit{(0.017)}                               & \textit{(3.679)}                       &                                &                                       &                      & \textit{(0.026)}              \\ \hline
\textbf{CL-RB}        & 0.840                                 & 3.668                         & -                              & 0.813                                 & 4.010                & -                    \\
\textbf{}             & \textit{(0.008)}                               & \textit{(0.338)}                       &                                & \textit{(0.008)}                               & \textit{(0.483)}              &                      \\ \hline
\textbf{NG-RB}        & 0.292                                 & 12.447                        & -                              & -                                     & -                    & 0.201                \\
\textbf{}             & \textit{(0.017)}                               & \textit{(3.003)}                       &                                &                                       &                      & \textit{(0.035)}             
\end{tabular}
\caption{Parameters estimates for the copula models chosen according to table ---add table--- and corresponding standard errors (in italic): $\rho$ is the correlation parameter and $\nu$ is the number of degrees of freedom of a Student-$t$ copula, while $\theta$ is the parameter of a Clayton copula. }
\label{tab:copulafit_parameters}
\end{table}

The correlation among pairs of assets involving NG is weak in both periods (in Period 1, the corresponding $\rho$ is lower than other pairs of assets, while in Period 2 the $\theta$ parameter is close to zero). In general, there seems to be a change in the structure of the dependence between the two periods. To better understand this change in the dependence structure between Period 1 and 2, we analyse several functionals of the dependence: the Spearman's $\rho_S$, the Kendall's $\tau$ and the tail dependence coefficients, $\lambda_L$ and $\lambda_U$, together with measure of distance for the independence copula, $\delta^2$ as defined in Equation \eqref{eq:delta} and $\mathcal{H}^2$ as defined in Equation \eqref{eq:h2}. 

\begin{footnotesize}
\begin{center}
  \begin{longtable}{l| c | c |c | c |c | c}
     & \textbf{$\rho_S$ } & \textbf{ $\tau$} & \textbf{$\lambda_L$} & \textbf{$\lambda_U$} & $\delta^2$ & $\mathcal{H}^2$\\
    \hline
    \multicolumn{7}{l}{\textbf{Period 1}} \\
    \hline
    BZ$-$CL	&	0.926	&	0.793	&	\multicolumn{2}{c|}{0.697} & 0.889 & 0.923  \\
        	&	\textit{(0.001)}    &   \textit{(0.008)}     &    \multicolumn{2}{c|}{\textit{(0.004)}}	& \textit{(0.001)}\\
    BZ$-$NG	&	0.3430	&	0.2343	&	\multicolumn{2}{c|}{0.0379}	& 0.174 & 0.351 \\
        	&	\textit{(0.005)}    &   \textit{(0.026)}    &   \multicolumn{2}{c|}{\textit{(0.042)}}	& \textit{(0.002)} \\
    BZ$-$RB	&	0.8163	&	0.6332	&	\multicolumn{2}{c|}{0.5716}	& 0.701	& 0.809\\
        	&	\textit{(0.001)}	&		\textit{(0.010)} &	\multicolumn{2}{c|}{\textit{(0.025)}}	& \textit{(0.005)}\\
    CL$-$NG	&	0.3316	&	0.2263	&	\multicolumn{2}{c|}{0.0375}		& 0.186 
    & 0.382 \\
        	&	\textit{(0.001)}	&	\textit{(0.027)}	&	\multicolumn{2}{c|}{\textit{(0.037)}}	& \textit{(0.010)} \\
    CL$-$RB	&	0.8145	&	0.6350	&	\multicolumn{2}{c|}{0.5959}	& 0.712 
    & 0.825 \\
        	&	\textit{(0.002)}	&	\textit{(0.012)}	&	\multicolumn{2}{c|}{\textit{(0.010)}}	& \textit{(0.012)} \\
    NG$-$RB	&	0.3187	&	0.2172	&	\multicolumn{2}{c|}{0.0368}	& 0.157	
    & 0.362 \\
        	&	\textit{(0.002)}	&	\textit{(0.027)}	&	\multicolumn{2}{c|}{\textit{(0.042)}}	& \textit{(0.016)} \\
    \hline
    \multicolumn{6}{l}{\textbf{Period 2}} \\
    \hline
    BZ$-$CL	&	0.8703 &	0.7055	&	\multicolumn{2}{c|}{0.4329} & 0.838	& 0.890\\
        	&	\textit{(0.013)}	&	\textit{(0.015)} & \multicolumn{2}{c|}{\textit{(0.162)}} & \textit{(0.016)}	\\
    BZ$-$NG    &   0.1178	&	0.0786	&	0.0172	& 0.0000 & 0.080 & 0.135\\
               & \textit{(0.001)}	&	\textit{(0.014)}	    &	\textit{(0.022)} & \textit{(0.000)}	
               & \textit{0.014} \\
    BZ$-$RB	&	0.8124	&	0.6354	&	\multicolumn{2}{c|}{0.6054}	& 0.741	
    & 0.818 \\
        	&	\textit{(0.001)}	&   \textit{0.013}	&	\multicolumn{2}{c|}{\textit{(0.010)}}	& \textit{0.008} \\
    CL$-$NG	&	0.1317	&	0.0880	&	0.0275 & 0.0000	 & 0.065	&
    0.147 \\
            &	\textit{(0.001)}	&	\textit{(0.013)}    &	\textit{(0.022)} & \textit{(0.000)}  & \textit{(0.024)}\\
    CL$-$RB	&	0.7340	&	0.5531	&	\multicolumn{2}{c|}{0.5226}	& 0.630	&
    0.760 \\
        	&	\textit{(0.000)}    &	\textit{(0.013)}	&	\multicolumn{2}{c|}{\textit{(0.020)}}	& \textit{(0.002)}\\
    NG$-$RB	&	0.1015	&	0.0676	&   0.0084	& 0.0000 & 0.065 & 0.075	\\
            &	\textit{(0.000)}	&	\textit{(0.019)}	&	\textit{(0.008)} & \textit{(0.000)} & \textit{(0.009)}	\\
    \caption{Functional estimates with corresponding standard errors in brackets: $\rho_S$ is the Spearman's $\rho$, $\tau$ is the Kendall's $\tau$, $\lambda_L$ and $\lambda_U$ are the lower and upper tail dependences indices respectively, $\delta^2$ is the mutual information index, $\mathcal{H}^2$ is the Hellinger correlation coefficient.}
    \label{tab:copula_functionals}
  \end{longtable}
\end{center}
\end{footnotesize}

Table \ref{tab:copula_functionals} shows that the monotonic dependence decreases from Period 1 to Period 2, for all the pairs of assets; in particular the dependence described by $\rho_S$ and $\tau$ becomes close to zero for the pair of assets including NG; it is interesting to notice that, while the dependence is small for the pairs involving NG, each functional is still significantly different from zero and this fact leads to reject the hypothesis of independence. 

Tail dependence is indicated by a single value in Period 1 because in the case of the Student-$t$ copula $\lambda_L=\lambda_U$. In Period 2, when the Clayton copula is selected, $\lambda_U=0$ by definition. From Period 1 to Period 2, tail dependence tend to decrease except that for the pair BZ-RB.

The two last columns of Table \ref{tab:copula_functionals} show the estimated mutual information index and the Hellinger coefficient, which are measures of discrepancy between the joint distribution and the case of independence. The indices decrease between the two periods under considerations (except for the pair BZ-RB) and it shows that the bivariate distribution of BZ-CL, BZ-RB and CL-RB seems to be far from the case of independence, while the bivariate distributions involving NG tend to show lower dependence, in particular in Period 2. 

The results of this section seem to suggest that the strength of dependence, both overall and in the tails of the joint distributions, decreases in periods where the occurrence of extraordinary OPEC meetings decreases.

\begin{figure}[h!]
  \centering
  \includegraphics[width=10cm,height=7cm]{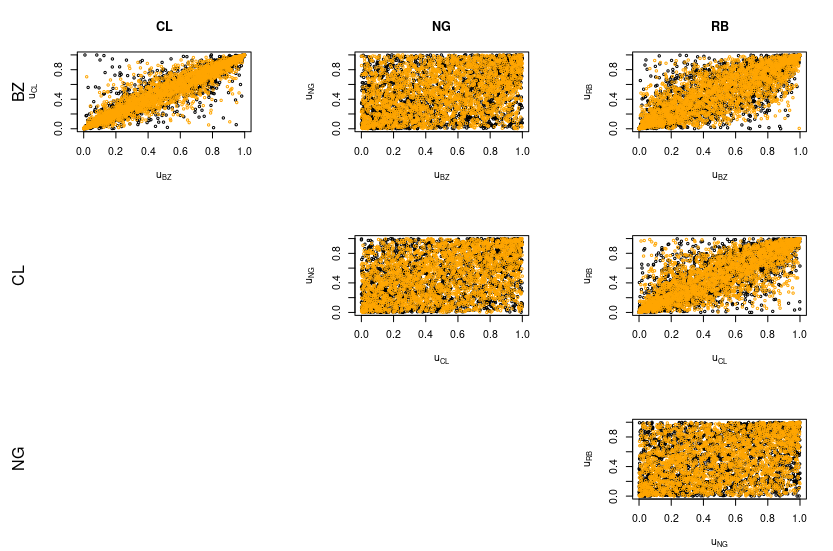}
  \caption{Dependence scatterplots for pseudo-observations (black) and observations simulated from the fitted models (orange) in Period 1.}
  \label{fig:scatter_cop1}
\end{figure}

\begin{figure}[h!]
  \centering
  \includegraphics[width=10cm,height=7cm]{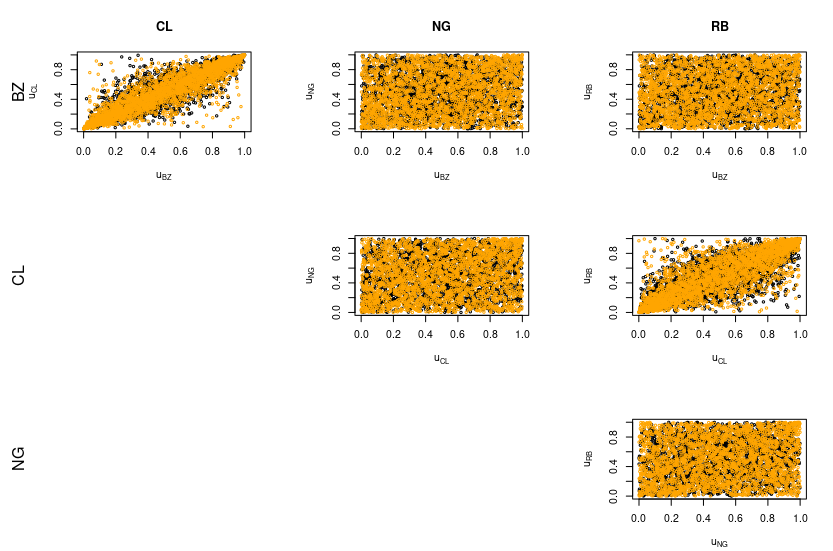}
  \caption{Dependence scatterplots for pseudo-observations (black) and observations simulated from the fitted models (orange) in Period 2.}
  \label{fig:scatter_cop2}
\end{figure}

Figure \ref{fig:scatter_cop1} and \ref{fig:scatter_cop2} show the scatterplots of the pseudo-observations obtained via nonparametric estimation of the marginals in Period 1 and 2, respectively. For a visual comparison between the observations and the fitted models, we have simulated new pseudo-data from the estimated models and plotted together with the original data. It is evident that CL and BZ show the strongest dependence, while RB shows an average level of dependence with both CL and BZ. As expected, NG has the lowest level of dependence with all the other assets. 
Moreover, the scatterplots show a visual change in the dependence structure, where assets in Period 1 seem to be more strongly dependent than in Period 2. Finally, the comparison with the simulated pesudo-observations (in orange in the plots) seem to support the choice of the models. 

\section{Discussion}
\label{sec:conclu}

We examined the dependence structure of crude oil prices in periods of irregular and regular OPEC announcements. We performed an analysis to check the best copula model to represent co-movements between crude oil benchmark prices and their derivatives. We found evidence of symmetric tail dependence, captured by a Student-$t$ copula for Brent crude oil and WTI crude oil, Brent crude oil and gasoline RB and WTI crude oil and gasoline RB. We saw a consistent decreasing of the dependence among asset prices in periods of regular OPEC announcements, regarding both the overall dependence (investigated through the Spearman's $\rho_S$ and the Kendall's $\tau$) and the tail dependence. In particular, our analysis showed that natural gas prices, which are generally less influenced by crude oil prices, show a dependence close to zero with the other asset prices we investigated in this work in the period of regular announcements. 

To the best of our knowledge, this analysis is novel in the literature: in times of crude oil markets stress, the co-movements of crude oil prices and derivatives are more strongly related; on the other hand, in periods of stable global demand of crude oil, OPEC decreases the number of extraordinary meetings and the dependence structure among crude oil and derivatives' prices is reduced. Interestingly, the only case where the level of tail dependence increases from the first to the second period is the pair of Brent crude oil and gasoline RB; while in all other combinations, tail dependence decreases. The proposed methodology has the important advantage to realistically model each asset price series separately, avoiding the need to implement simplistic joint distributions. Moreover, several assumptions on the type of the dependence can be tested and the best model for the co-movements can be chosen. While we do not aim at establishing a causal relationship between the regularity of the OPEC announcements and the strenght of the dependence, we think that the findings have important implications for risk management: in period of stress, tail dependence tends to be higher, i.e. joint losses in crude oil markets are more likely; on the other hand, it tends to decrease in periods where OPEC does not need to adjust prices through output. 


%
%
%
%


\section*{Acknowledgements}

This research did not receive any specific grant from funding agencies in the public, commercial, or not-for-profit sectors.

\newpage





\end{document}